%% file: main.tex
\documentclass[11pt,oneside]{article}
\input{config}
\newcommand{\blind}{1}

\def\bbR{{\mathbb{R}}}

\newcommand\mytext[1]{\text{\scriptsize{#1}}}

\allowdisplaybreaks

\begin{document}
\setlength{\abovedisplayskip}{5pt}
\setlength{\belowdisplayskip}{5pt}

\if1\blind
{
  \begin{center}
{\LARGE\bf Distributionally Enhanced Marginal Sensitivity Model and Bounds}\\
\vspace{.1in}
Yi Zhang\footnotemark[1], Wenfu Xu\footnotemark[2], Zhiqiang Tan\footnotemark[1]\\
\footnotetext[1]{Department of Statistics, Rutgers University}
\footnotetext[2]{College of Economics and Management, China Jiliang University}
\vspace{.1in}
\today
\end{center}
} \fi

\if0\blind
{
  \bigskip
  \bigskip
  \bigskip
  \begin{center}
    {\LARGE\bf Distributionally Enhanced Marginal Sensitivity Model and Bounds}
\end{center}
  \medskip
} \fi

\bigskip
\begin{abstract}
For sensitivity analysis against unmeasured confounding, we build on the marginal sensitivity model (MSM) and propose a new model, deMSM, by incorporating a second constraint on the shift of potential outcome distributions caused by unmeasured confounders in addition to the constraint on the shift of treatment probabilities.
We show that deMSM leads to interpretable sharp bounds of common causal parameters and tightens the corresponding MSM bounds. Moreover, the sharp bounds are symmetric in the two deMSM constraints, which facilitates practical applications. Lastly, we compare deMSM with other MSM-related models in both model constraints and sharp bounds, and reveal new interpretations for later models.
\end{abstract}

\noindent%
\vfill











\section{Introduction}\label{sec:intro}
\input{Sec1Intro}

\vspace{-.5em}
\section{Notations and related models}\label{sec:setup}
\input{Sec2sens}

\section{Proposed sensitivity model and bounds}\label{sec:propsens}
\input{Sec3deMSM}

\section{Implications about eMSM}\label{sec:implication}
\input{Sec4Implication}

\section*{Acknowledgements}
Yi Zhang and Zhiqiang Tan are partially supported by NIH grant 1R01LM014257.

\bibliographystyle{apalike}
\bibliography{references}
\newpage
\input{supplementary}
\bibliographystyleappend{apalike}
\bibliographyappend{references.bib}


\end{document}

%% file: config.tex
\usepackage{epsfig,lscape,amsmath, amssymb,amsfonts, amsthm, bm, bbm, graphicx,geometry,float, subcaption, enumerate, color, verbatim ,mathtools,algorithm, algorithmic, url, amssymb, bm,setspace, natbib}
\usepackage[svgnames,x11names,table]{xcolor}
\usepackage[colorlinks=true,citecolor=Navy,linkcolor=.,urlcolor=black]{hyperref}
\mathtoolsset{showonlyrefs}

\usepackage{tikz}
\usetikzlibrary{arrows.meta, positioning}
\usepackage{setspace}
\usepackage[utf8]{inputenc}
\usepackage[english]{babel}
\usepackage{framed}
\usepackage{multibib}
\usepackage[symbol]{footmisc}
\usepackage{scalerel}
\newcites{append}{Supplement References}

\newtheorem{pro}{Proposition}

\theoremstyle{definition}

\theoremstyle{definition}

\usepackage[medium,compact]{titlesec}

\titlespacing*{\section} {0pt}{1.5ex}{1ex}
\titlespacing*{\subsection} {0pt}{1.5ex}{1ex}
\titlespacing*{\subsubsection} {0pt}{1ex}{1ex}

\graphicspath{{figures/}}
\allowdisplaybreaks

\newcommand{\ud}{\mathrm{d}}

\newcommand{\odds}{\mathrm{odds}}

\theoremstyle{definition}

\newcommand{\tauG}{\tau_{\scaleto{\Gamma}{4pt}}}

%% file: Sec1Intro.tex
For causal inference, we consider the potential-outcome framework \citep{rubin1974estimating, Neyman1923}.
Let $X$ be a vector of measured covariates, let $T$ be a binary treatment taking value $0$ or $1$,
and let $Y^0$ and $Y^1$ be the potential outcomes associated with treatment 0 and 1 respectively.
The observed outcome $Y$ is determined from one of the potential outcomes, depending on $T$,
through a consistency assumption, $Y=TY^1+(1-T)Y^0$.
At the unit level, the observed data consists of $(Y,T,X)$, and the potential outcomes $(Y^0,Y^1)$ can be viewed as latent variables.

A key task of causal inference is to infer population parameters about $(Y^0,Y^1)$, such as
the means, $\mu^0 = E(Y^0)$ and $\mu^1 = E(Y^1)$, and the average treatment effect (ATE), $\mu^1-\mu^0$.
Point identification of these parameters typically relies on the unconfoundedness assumption:
$(Y^0,Y^1)\perp T|X$, that is, $(Y^0,Y^1)$ is conditionally independent of $T$ given $X$.
However, this assumption is untestable and may often be violated due to unmeasured confounding.
For sensitivity analysis, a sensitivity model is used to quantify unmeasured confounding and
then the causal parameter of interest can be partially identified or bounded.
There is an extensive and growing literature on sensitivity analysis.
See, for example, \citet{PRosenbaum2002}, \citet{tan2006distributional, tan2024model}, and references therein.

In this paper, we propose a new sensitivity model that extends the marginal sensitivity model (MSM) of \citet{tan2006distributional}.
The new model is called the distributionally enhanced MSM (deMSM), which incorporates an outcome sensitivity constraint in the form of density ratio bounds (hence a distributional constraint) for potential outcomes, in addition to the treatment sensitivity constraint as found in a variation of MSM with an unmeasured confounder \citep{dorn2023sharp}. We study the sharp bounds of common causal parameters under deMSM,
and show that deMSM sharp bounds are readily interpretable and, remarkably, symmetric in the treatment and outcome sensitivity parameters. We also compare and connect deMSM with other MSM-related models \citep{dorn2023sharp,zhang2025enhanced}.
We find that (i) the deMSM sharp bounds are always tighter than the corresponding MSM bounds with only the treatment sensitivity constraint, and
(ii) the sharp bounds from deMSM can be made to match those from enhanced MSM (eMSM) with the recommended specification in \cite{zhang2025enhanced}, thereby allowing the outcome sensitivity parameter in eMSM to be re-interpreted via outcome density ratios as in deMSM.

%% file: Sec2Sens.tex
We focus on the inference about $\mu^1$ and defer results on $\mu^0$ and ATE to the Supplement.
The marginal sensitivity model (MSM) of \citet{tan2006distributional} characterizes unmeasured confounding via the Radon--Nikodym derivative (or density ratio) between two conditional distributions of $Y^1$:
\begin{align}\label{eq:msm-2006}
    \lambda^*_1(Y^1,X)=\frac{\ud P(Y^1|T=0,X)}{\ud P(Y^1|T=1,X)}\in [\Lambda_1(X),\Lambda_2(X)],
\end{align}
where the sensitivity parameters $\Lambda_2 (X) \ge 1 \ge \Lambda_1(X) \ge 0$ are \textit{pre-specified} covariate functions.

A variation of MSM \citep{dorn2023sharp} introduces an unmeasured covariate $U$ (as a latent variable) accounting for all unmeasured confounding such that
\begin{equation}
    (Y^0,Y^1)\perp T | X,U, \label{eq:unm-conf-indep}
\end{equation}
and bounds the distributional ratio of $U$ instead of $Y^1$:
\begin{align}
 \lambda^*(X,U)=\frac{\ud P(U|T=0,X)}{\ud P(U|T=1,X)}=\frac{P(T=1|X)}{P(T=0|X)}\Big/\frac{P(T=1|X,U)}{P(T=0|X,U)} \in [\Lambda_1(X),\Lambda_2(X)], \label{eq:msm}
\end{align}
where the odds ratio expression of $\lambda^*$ follows from Bayes' rule, and $(\Lambda_1,\Lambda_2)$ are the same as in \eqref{eq:msm-2006}.
To distinguish from MSM defined by \eqref{eq:msm-2006}, we refer to \eqref{eq:unm-conf-indep}--\eqref{eq:msm} as MSM-U. MSM and MSM-U induce the same sharp bounds of $\mu^1$. See \citet{zhang2025enhanced} for additional discussions of the two and their equivalent models.
For both models, $\Lambda_1(X) = \Lambda_2(X) \equiv 1$ indicates unconfoundedness. The further $\Lambda_1(X)$ and $\Lambda_2(X)$ are away from $1$, the greater unmeasured confounding is allowed.

A potential limitation of MSM-U(although not apparent for MSM) is that it only constrains the association between $U$ and treatment assignment $T$. However, to induce unmeasured confounding, $U$ needs to be associated with both treatment assignment $T$ and outcome $(Y^0, Y^1)|T$ conditionally on $X$ (e.g., \cite{greenland1986identifiability}).
Based on this observation, \citet{zhang2025enhanced} proposed an enhanced marginal sensitivity model (eMSM) that complements the treatment sensitivity constraint \eqref{eq:msm} with the mean outcome sensitivity constraint
\begin{align}\label{eq:emsm}
     \eta^*(X,U) - E(Y^1|T=1,X)\in[-\Delta_1(X), \Delta_2(X)],
\end{align}
where $\eta^*(X,U)=E(Y^1|X,U)=E(Y|T=1,X,U)$, and the outcome sensitivity parameters $\Delta_1(X), \Delta_2(X) \ge 0$ are \textit{pre-specified} covariate functions. When $\Delta_1(X)=\Delta_2(X)\equiv0$, $\eta^*(X,U)=E(Y|T=1,X)$, and $\mu^1$ equals its reference value under unconfoundedness, $E\{E(Y|T=1,X)\}$. The larger $\Delta_1(X)$ and $\Delta_2(X)$ are, the more $U$ may influence $Y^1$ given $T=1$ and $X$, and hence the greater unmeasured confounding is allowed. Both constraints \eqref{eq:msm} and \eqref{eq:emsm} are ``marginal'' with respect to how the probabilities of $T$ given $(X,U)$ and the mean of $Y^1$ given $(T=1,X,U)$ may differ from those with $U$ marginalized.

We take a frequentist perspective and denote by $P$ the true distribution of the full data $(Y^0,Y^1,T,X,U)$ as done above.
The induced distribution of $P$ on $(Y,T,X)$ is the true distribution of the observed data.
The expectation with respect to $P$ is denoted by $E(\cdot)$.
Conceptually, we identify a sensitivity model with the collection of joint distributions of $(Y^0,Y^1,T,X,U)$ such that for each individual distribution $Q$
\begin{itemize}\addtolength{\itemsep}{-.1in}
\item[(i)] $Q$ is compatible with the observed-data distribution: $Q$ and $P$ induce the same distribution of $(Y,T,X)$ under the consistency assumption;
\item[(ii)] $Q$ is feasible in the sensitivity model, e.g., in the case of eMSM, properties \eqref{eq:unm-conf-indep}, \eqref{eq:msm} and \eqref{eq:emsm} hold under $Q$ with $(\lambda^*,\eta^*)$ replaced by their $Q$-analogue $(\lambda_Q,\eta_Q)$.
\end{itemize}
Then a sensitivity model is correct if the true distribution $P$ is contained in the associated collection of joint distributions. We use an asterisk in notations, such as $\pi^*$ and $\lambda^*$, to signify quantities induced from the true distribution $P$ as opposed to a general distribution $Q$.

From the above perspective, the sharp upper bound of $\nu^1(X)=E(Y^1|T=0,X)$ and that of $\mu^1$ under a sensitivity model are defined as
\begin{align}
\begin{split}
    &\nu^{1+}(X)=\sup_Q\;E_Q (Y^1|T=0,X)\\
    &\mu^{1+}=\sup_Q \;E_Q (Y^1)=E \left\{TY+(1-T) \nu^{1+}(X) \right\},\label{eq:sharp-mu1}
\end{split}
\end{align}
where $Q$ satisfies (i) and (ii), and $E_Q(\cdot)$ denotes the expectation with respect to $Q$. Sharp lower bounds are similarly defined with infimum replacing supremum in \eqref{eq:sharp-mu1}.

\citet{zhang2025enhanced} shows that the eMSM sharp upper bound of $\nu^{1}(X)$ and that of $\mu^{1}$ are
\begin{align}\label{eq:eMSM-mu1-upper}
\begin{split}
\nu^{1+}_{\mytext{eMSM}} (X)&= E(Y|T=1,X)+\{\Lambda_2(X)-\Lambda_1(X)\}\times\\
&\hspace{1em}\min\big[\tau(X) \Delta_1(X), \{1-\tau(X)\}\Delta_2(X),E \{\rho_{\tau}(Y,q^*_{1,\tau})|T=1,X \}\big],\\
\mu^{1+}_{\mytext{eMSM}}&=E \left\{TY+(1-T) \nu^{1+}_{\mytext{eMSM}}(X) \right\},
\end{split}
\end{align}
where $\tau=\tau(X)=\{\Lambda_2(X)-1\}/\{\Lambda_2(X)-\Lambda_1(X)\}$ and is set to $1/2$ when $\Lambda_1(X)=\Lambda_2(X)=1$, $q^*_{1,\tau}=q^*_{1,\tau}(X)$ is the $\tau(X)$-quantile of $Y| T=1,X$ defined as a minimizer of the quantile loss $E \{\rho_{\tau} (Y, q) | T=1,X\}$ over $q\in\bbR$ \citep{koenker1978regression}, $\rho_{\tau} (y,q)=  \tau (y-q)_{+}+(1-\tau) (q-y)_{+}$ is the check function, and $c_+=\max\{c,0\}$ for $c \in \bbR$.

The eMSM sharp bounds are related to the MSM ones. For brevity, we use $\rho_{1,\tau}^*$ as a shorthand for $\rho_{\tau}(Y,q^*_{1,\tau})$. When $\Delta_1$ and $\Delta_2$ are large so that $\min[\tau(X) \Delta_1(X), \{1-\tau(X)\}\Delta_2(X)]\ge E(\rho_{1,\tau}^*|T=1,X )$, $\nu^{1+}_{\mytext{eMSM}} (X)$ reduces to
\begin{align}\label{eq:msm-sharp}
    \nu^{1+}_{\mytext{MSM}} (X)=E(Y|T=1,X)+\{\Lambda_2(X)-\Lambda_1(X)\}E\left(\rho_{1,\tau}^*|T=1,X \right),
\end{align}
which is the sharp upper bound of $\nu^1(X)$ under MSM and under MSM-U \citep{dorn2023sharp, tan2024model}. In this case, $\mu^{1+}_{\mytext{eMSM}}$ reduces to $\mu^{1+}_{\mytext{MSM}}$. When $\Delta_1$ and $\Delta_2$ are small with respect to $E(\rho_{1,\tau}^*|T=1,X )$, the eMSM sharp bounds are tighter than their MSM counterparts.  

%% file: Sec3deMSM.tex
\subsection{Distributionally enhanced marginal sensitivity model}\label{sec:deMSM}

\begin{figure}
    \centering
    \begin{tikzpicture}[
        roundnode/.style={circle, draw=black, thick, minimum size=.6cm},
        directed edge/.style={-{Latex[round]}, thick},
        blue edge/.style={-{Latex[round]}, thick,dashed, black},
        dashed blue edge/.style={-{Latex[round]}, thick, dotted, black},
        edgelabel/.style={midway, font=\scriptsize, text=black},
        node distance=2cm
    ]

    \node[roundnode] (T1) at (0, 0) {T};
    \node[roundnode] (Y1) at (2.5, 0) {Y};
    \node[roundnode] (U1) at (1.15, 2) {U};

    \draw[blue edge] (U1) -- (T1) node[edgelabel,midway, left, xshift=2em] {\parbox{3cm}{Treatment \\constraint \eqref{eq:msm}}};
    \draw[blue edge] (U1) -- (Y1) node[edgelabel,midway, right, xshift=2pt] {\parbox{3cm}{Distributional outcome\\ constraint \eqref{eq:demsm}}};
    \draw[directed edge] (T1) -- (Y1);

    \node[roundnode] (T2) at (7, 0) {T};
    \node[roundnode] (Y2) at (9.5, 0) {Y};
    \node[roundnode] (U2) at (8.15, 2) {U};

    \draw[blue edge] (U2) -- (T2) node[edgelabel,midway, left, xshift=3em] {\parbox{3cm}{ Treatment \\constraint \eqref{eq:msm}}};
    \draw[blue edge] (U2) -- (Y2) node[edgelabel,midway, right, xshift=2pt] {\parbox{3cm}{Mean outcome\\ constraint \eqref{eq:emsm}}};
    \draw[directed edge] (T2) -- (Y2);

    \end{tikzpicture}
    \caption{Graphical representations of deMSM (left) and eMSM (right) constraints conditional on $X$.} 
    \label{fig:dag-compare}
\end{figure}
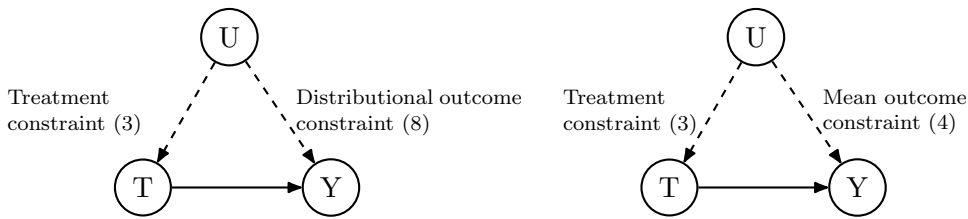

While eMSM provides tighter sharp bounds than MSM and MSM-U, sensitivity analysis based on eMSM requires more care due to
additional sensitivity parameters $(\Delta_1,\Delta_2)$.
One complication is that the interpretation of $(\Delta_1,\Delta_2)$ is data-dependent, i.e.,
the meanings of $(\Delta_1,\Delta_2)$ are not comparable between different studies (i.e., between different observed-data distributions).
\citet{zhang2025enhanced} tackled this issue by employing a re-parameterization of $(\Delta_1,\Delta_2)$ with a data-independent sensitivity parameter $\delta$.
See Section \ref{sec:implication} for further discussion.
Another, more fundamental issue is that the effect of $U$ on $T$ is measured by the density ratio as in \eqref{eq:msm}, while that of $U$ on $Y^1$ is measured by the difference in expectations as in \eqref{eq:emsm}. The two measurements are on different scales, making it difficult to
assess the overall strength of unmeasured confounding.

To improve interpretability while maintaining constraints on the effect of $U$ on $Y^1$ as in eMSM, we propose the distributionally enhanced marginal sensitivity model (deMSM). For inference about $\mu^1$, a deMSM assumes, in addition to \eqref{eq:msm}, a distributional outcome sensitivity constraint:
\begin{align}\label{eq:demsm}
    \omega^*(Y^1,X,U)=\frac{\ud P(Y^1|X,U)}{\ud P(Y^1|T=1,X)}\in[\Gamma_1(X),\Gamma_2(X)],
\end{align}
where the outcome sensitivity parameters $\Gamma_2(X)\ge1\ge\Gamma_1(X)\ge0$ are \textit{pre-specified}.
When $\Gamma_2(X)=\Gamma_1(X)\equiv1$, $Y^1\perp U|(T=1,X)$ and $\mu^1$ is identified by its reference value under unconfoundedness, $E\{E(Y|T=1,X)\}$. The further $\Gamma_1(X)$ and $\Gamma_2(X)$ are away from $1$, the greater unmeasured confounding is allowed.
The new outcome sensitivity constraint \eqref{eq:demsm} is ``marginal'' in the same sense as the constraints \eqref{eq:msm} and \eqref{eq:emsm}.

Compared with the mean constraint \eqref{eq:emsm} in eMSM, \eqref{eq:demsm} is a distributional constraint as density ratio bounds. The difference is illustrated in the two directed acyclic graphs in Figure \ref{fig:dag-compare}.
Hence in contrast with $(\Delta_1,\Delta_2)$ in eMSM, the sensitivity parameters $(\Gamma_1,\Gamma_2)$ are on the scale of density ratios similarly to $(\Lambda_1,\Lambda_2)$.
More remarkably, as shown by Proposition~\ref{pro:demsm-mu1-up} later, the sharp bounds of $\mu^1$ are symmetric in the two pairs of parameters under deMSM.

We comment on the relation between deMSM and MSM.
First, the two deMSM constraints \eqref{eq:msm} and \eqref{eq:demsm} together imply an MSM constraint:
\begin{align}
\bar\Lambda_1(X)\le\lambda^*_1(Y^1,X)\le\bar\Lambda_2(X)\label{eq:implied-msm2006},
\end{align}
with the \emph{implied} MSM sensitivity parameters
\begin{align}
&\bar\Lambda_1(X)=1 -\{\Lambda_2(X)-\Lambda_1(X)\}\{\Gamma_2(X)-\Gamma_1(X)\} \min\big[ \{1-\tau(X)\}\{1-\tau_{\scaleto{\Gamma}{4pt}}(X)\}, \tau(X)\tau_{\scaleto{\Gamma}{4pt}}(X) \big],\\
    &\bar\Lambda_2(X)=1 +\{\Lambda_2(X)-\Lambda_1(X)\}\{\Gamma_2(X)-\Gamma_1(X)\} \min\big[ \{1-\tau(X)\}\tau_{\scaleto{\Gamma}{4pt}}(X), \tau(X)\{1-\tau_{\scaleto{\Gamma}{4pt}}(X)\} \big],
\end{align}
where $\tauG$ is defined the same as $\tau$ with $(\Gamma_1,\Gamma_2)$ replacing $(\Lambda_1,\Lambda_2)$.
Consequently, deMSM with sensitivity parameters $(\Lambda_1,\Lambda_2)$ and $(\Gamma_1,\Gamma_2)$ is a submodel of the implied MSM with parameters $(\bar\Lambda_1,\bar\Lambda_2)$
, i.e., any distribution $Q$ that belongs to the deMSM also belongs to the implied MSM.
Second, the implied sensitivity parameters $(\bar\Lambda_1,\bar\Lambda_2)$ are symmetric in $(\Lambda_1,\Lambda_2)$ and $(\Gamma_1,\Gamma_2)$
from the constraint \eqref{eq:msm} and \eqref{eq:demsm} respectively, and satisfy
\begin{align}\label{eq:impliedpara-demsmpara}
   \max\{\Lambda_1(X),\Gamma_1(X)\}\le \bar\Lambda_1(X)\le1\le \bar\Lambda_2(X)\le\min\{\Lambda_2(X),\Gamma_2(X)\}.
\end{align}
Finally, by the two comments above, the sharp bounds from deMSM are at least as tight as from the implied MSM, which are in turn at least as tight as from MSM
with parameters $(\max\{\Lambda_1,\Gamma_1\}, \min\{\Lambda_2,\Gamma_2\})$.
See Supplement Section \ref{sec:proof-implied-msm2006} for derivations of \eqref{eq:implied-msm2006} and \eqref{eq:impliedpara-demsmpara}.

\subsection{Population sharp bounds} \label{sec:sharpbounds}
Under deMSM defined by \eqref{eq:unm-conf-indep}, \eqref{eq:msm} and \eqref{eq:demsm}, the sharp lower and upper bounds of $\mu^1$, as in \eqref{eq:sharp-mu1}, are
\begin{align}
\begin{split}\label{eq:demsm-sharpbds-mu1}
    & \mu^{1-}_{\mytext{deMSM}} = \inf_Q\;E_Q \left(Y^1 \right)= E \left\{TY+(1-T) \nu^{1-}_{\mytext{deMSM}}(X) \right\}, \\
&\mu^{1+}_{\mytext{deMSM}} = \sup_Q \;E_Q \left(Y^1 \right)=E \left\{TY+(1-T) \nu^{1+}_{\mytext{deMSM}}(X) \right\} ,
\end{split}
\end{align}
where the optimizations are over $Q$, a joint distribution of $(Y^0,Y^1,T,X,U)$, allowed in deMSM, and $\nu^{1-}_{\mytext{deMSM}}(X)=\inf_QE_Q (Y^1|T=0,X)$ and $\nu^{1+}_{\mytext{deMSM}}(X)=\sup_QE_Q (Y^1|T=0,X)$ over $Q$ as in \eqref{eq:demsm-sharpbds-mu1} are the sharp lower and upper bounds of $\nu^1(X)$ under the deMSM. The deMSM sharp bounds are solved as follows. See the proof in Supplement Section \ref{sec:proof-prop1} for the construction of a joint distribution $Q$ that attains the sharp bounds.

\begin{pro}[Sharp bounds]\label{pro:demsm-mu1-up}
The sharp lower bound and upper bound for
$\nu^1=E(Y^1|T=0)$ are respectively
\begin{align}
\begin{split}
    &\nu^{1-}_{\mytext{deMSM}}(X) = E(Y|T=1,X)-\left\{\bar\Lambda_2(X)-\bar\Lambda_1(X)\right\}E \left(\rho_{1,1-\bar\tau}^*|T=1,X \right),\\
&\nu^{1+}_{\mytext{deMSM}}(X) = E(Y|T=1,X)+\left\{\bar\Lambda_2(X)-\bar\Lambda_1(X)\right\}E \left(\rho_{1,\bar\tau}^*|T=1,X \right).\label{eq:demsm-nu1-up}
\end{split}
\end{align}
The sharp lower bound and upper bound for $\mu^1=E(Y^1)$ are $\mu^{1-}_{\mytext{deMSM}} = E \{TY+(1-T) \nu^{1-}_{\mytext{deMSM}} \}$ and $\mu^{1+}_{\mytext{deMSM}} = E \{TY+(1-T) \nu^{1+}_{\mytext{deMSM}}\}$ respectively. Here $\rho_{1,\gamma}^*=\rho_{\gamma}(Y,q^*_{1,\gamma})$ for $\gamma=\bar\tau$ or $1-\bar\tau$, $\bar\tau(X)=\{\bar\Lambda_2(X)-1\}/\{\bar\Lambda_2(X)-\bar\Lambda_1(X)\}$ and is set to $1/2$ if $\bar\Lambda_1(X)=\bar\Lambda_2(X)=1$, and $(\bar\Lambda_1(X),\bar\Lambda_2(X))$ are the implied sensitivity parameters in \eqref{eq:implied-msm2006}.
\end{pro}

We provide some remarks about the deMSM sharp bound $\nu^{1+}_{\mytext{deMSM}}(X)$. Analogous statements on $\nu^{1-}_{\mytext{deMSM}}(X)$, $\mu^{1-}_{\mytext{deMSM}}$, and $\mu^{1+}_{\mytext{deMSM}}$ also hold. To start, it is helpful to express \eqref{eq:demsm-nu1-up} in the sensitivity parameters $(\Lambda_1,\Lambda_2)$ and $(\Gamma_1,\Gamma_2)$ directly: 
\begin{align}
\begin{split}\label{eq:nu1+-min}
    &\nu^{1+}_{\mytext{deMSM}} (X) = E(Y|T=1)+(\Lambda_2-\Lambda_1)(\Gamma_2-\Gamma_1)\times\\
   &\min\big\{\tau_{\scaleto{\Gamma}{4pt}}E \left(\rho_{1,1-\tau}^*|T=1\right),(1-\tau_{\scaleto{\Gamma}{4pt}}) E \left(\rho_{1,\tau}^*|T=1\right),\tau E \left(\rho_{1,1-\tau_{\scaleto{\Gamma}{3pt}}}^*|T=1\right),(1-\tau)E \left(\rho_{1,\tau_{\scaleto{\Gamma}{3pt}}}^*|T=1\right)\big\},
\end{split}
\end{align}
where $X$ is made implicit in \eqref{eq:nu1+-min} for presentation simplicity.

First, the expression \eqref{eq:demsm-nu1-up} shows that $\nu^{1+}_{\mytext{deMSM}}(X)$ coincides with the MSM sharp upper bound of $\nu^1(X)$ with the implied sensitivity parameters $(\bar\Lambda_1,\bar\Lambda_2)$. This result sharpens the last comment in Section \ref{sec:deMSM} that the latter bound is a valid but possibly relaxed upper bound of $\nu^{1}(X)$ under the deMSM. The sharpness is established by constructing a joint distribution $Q$, as in Supplement Section \ref{sec:proof-prop1}, that attains \eqref{eq:demsm-nu1-up} and belongs to the deMSM collection of joint distributions.

Second, when $\Gamma_1(X)=0$ and $\Gamma_2(X)=+\infty$, $\nu^{1+}_{\mytext{deMSM}}(X)$ reduces to the MSM sharp upper bound $\nu^{1+}_{\rm{MSM}}(X)$ in \eqref{eq:msm-sharp}. Similarly, when $\Lambda_1(X)=0$ and $\Lambda_2(X)=+\infty$,
\begin{align}\label{eq:nu1+-demsm-outcome}
    \nu^{1+}_{\rm{deMSM}}(X)=\nu^{1+}_{\rm{d}}(X)=E(Y|T=1,X)+\left\{\Gamma_2(X)-\Gamma_1(X)\right\}E \left(\rho_{1,\tau_{\scaleto{\Gamma}{3pt}}}^*|T=1,X\right),
\end{align}
where $\nu^{1+}_{\rm{d}}(X)$ is the sharp upper bound of $\nu^{1}(X)$ under the model specified by \eqref{eq:demsm} alone. Interestingly, $\nu^{1+}_{\rm{d}}(X)$ coincides with $\nu^{1+}_{\mytext{MSM}}(X)$ with $(\Gamma_1,\Gamma_2)$ replacing $(\Lambda_1,\Lambda_2)$.
In view of \eqref{eq:impliedpara-demsmpara}, $\nu^{1+}_{\mytext{deMSM}}(X)$ is a tighter upper bound of $\nu^1(X)$ than both $\nu^{1+}_{\mytext{MSM}}(X)$ and $\nu^{1+}_{\rm{d}}(X)$ and any MSM sharp upper bound with $(\Lambda_1,\Lambda_2)$-parameters
looser than $(\max\{\Lambda_1,\Gamma_1\}, \min\{\Lambda_2,\Gamma_2\})$.


Third, the expression \eqref{eq:nu1+-min} indicates that the confounding strength measured by constraint \eqref{eq:msm} and that by \eqref{eq:demsm} are symmetric and interchangeable. When constraint \eqref{eq:msm} or \eqref{eq:demsm} is imposed alone with the same choice of sensitivity parameters, $(\Lambda_1(X), \Lambda_2(X)) =(\Gamma_1(X),\Gamma_2(X))$, the sharp upper bound of $\nu^1(X)$ remains the same (by the second point above). When both are imposed, switching $(\Lambda_1(X),\Lambda_2(X))$ and $(\Gamma_1(X),\Gamma_2(X))$ does not affect the sharp upper bound of $\nu^1(X)$. This property stems from the symmetry of the expression \eqref{eq:nu1+-min} as well as that of $(\bar\Lambda_1,\bar\Lambda_2)$
in the treatment sensitivity parameters $(\Lambda_1(X),\Lambda_2(X))$ and the outcome sensitivity parameters $(\Gamma_1(X),\Gamma_2(X))$. The symmetry is conceptually appealing and motivates the simple choice of specifying $\Lambda=\Gamma$ for $\Lambda_1^{-1} = \Lambda_2 = \Lambda$ and $\Gamma_1^{-1} = \Gamma_2 =\Gamma$, as discussed later in Section \ref{sec:demsm-application}.

Proposition \ref{pro:demsm-mu1-up} can be generalized to contrasts of $\mu^1$ and $\mu^0$, such as ATE. 
The deMSM sharp bounds of those parameters are identical to the MSM sharp bounds with the implied sensitivity parameters. This generalization results from the simultaneous sharpness that the sharp upper (lower) bound of $\mu^{1}$ and the sharp lower (upper) bound of $\mu^{0}$ under deMSM can be simultaneously attained by a single distribution $Q$. See Supplement Sections \ref{sec:deMSM-bounds} and \ref{sec:proof-prop1} for details.

\subsection{Considerations in application of deMSM}\label{sec:demsm-application}

In practice, sensitivity analysis often involves varying the magnitudes of the sensitivity parameters and examining the changes in causal estimates. It is thus convenient to specify constant sensitivity parameters. We follow the symmetric choice of \citet{tan2006distributional} by setting $\Lambda_1^{-1}(X)=\Lambda_2(X)\equiv\Lambda$ and $\Gamma_1^{-1}(X)=\Gamma_2(X)\equiv\Gamma$ for some constants $\Lambda,\Gamma\ge1$.

Sensitivity analysis using deMSM with constant $\Lambda$ and $\Gamma$ has a straightforward and informative interpretation,
similar to the interpretation of MSM \citep{tan2024model} but with the implied MSM sensitivity parameters $(\bar\Lambda_1,\bar\Lambda_2)$.
Specifically, by Proposition \ref{pro:demsm-mu1-up},
the sharp upper bound of $\mu^1$ deviates from its reference value under unconfoundedness, $E\{E(Y|T=1,X)\}$, by
\begin{align}
    (\bar\Lambda_2-\bar\Lambda_1)E\left\{(1-T)E \left(\rho_{1,\bar\tau}^*|T=1,X \right)\right\}.  \label{eq:deMSM-deviation}
\end{align}
This deviation is characterized by $(\bar\Lambda_1,\bar\Lambda_2)$ through the quantile level $\bar\tau$ and the implied range of confounding strength $(\bar\Lambda_2-\bar\Lambda_1)$.
When $\Lambda\le\Gamma$, the implied parameters $(\bar\Lambda_1,\bar\Lambda_2)$ simplifies to
\begin{align}\label{eq:para-shrink-Lam<Gam}
    &\bar\Lambda_1=1-(1-\Gamma_1)(1-\Lambda_1)=1-(1-\Gamma^{-1})(1-\Lambda^{-1}),\\ &\bar\Lambda_2=1+(1-\Gamma_1)(\Lambda_2-1)=1+(1-\Gamma^{-1})(\Lambda-1),
\end{align}
from which the implied quantile-level $\bar\tau=\tau$ and the implied range $\bar\Lambda_2-\bar\Lambda_1=(1-\Gamma^{-1})(\Lambda-\Lambda^{-1})$. Similarly, when $\Gamma\le\Lambda$, we have $\bar\Lambda_1=1-(1-\Lambda^{-1})(1-\Gamma^{-1})$ and $\bar\Lambda_2=1+(1-\Lambda^{-1})(\Gamma-1)$,
from which $\bar\tau=\tauG$ and $\bar\Lambda_2-\bar\Lambda_1=(1-\Lambda^{-1})(\Gamma-\Gamma^{-1})$.
Therefore, for the sharp upper bound, the smaller value between $\Lambda$ and $\Gamma$ identifies a ``base'' constraint that determines the quantile-level of sensitivity ($\tau$ or $\tau_\Gamma$), while the larger value determines the shrinkage level ($1-\Gamma^{-1}$ or $1-\Lambda^{-1}$ respectively)
for the range of confounding strength from the base constraint. The above statement also holds for the sharp lower bounds under deMSM.

Compared to MSM with one parameter $\Lambda$ specifying a quantile level, deMSM has the advantage that the two parameters $(\Lambda,\Gamma)$
determine both a quantile level and a shrinkage level.
To facilitate practical applications, we suggest setting $\Lambda=\Gamma$ (hence back to one sensitivity parameter).
For example, when $\Lambda=\Gamma=2$, the quantile level is $\bar\tau=2/3$, indicating 67\%-quantile for the upper bound or 33\%-quantile for the lower bound,
and the shrinkage level is $1/2$, indicating that the deviation \eqref{eq:deMSM-deviation} of the eMSM sharp upper (or lower) bound
is $1/2$ of the MSM sharp upper (or lower) bound from the unconfoundedness reference value.
The choice $\Lambda=\Gamma$ is motivated
by the symmetry and exchangeability of $\Lambda$ and $\Gamma$ (the third comment for Proposition \ref{pro:demsm-mu1-up})
and the monotonicity of deMSM sharp bounds in $\Lambda$ and $\Gamma$. The monotonicity indicates that a deMSM sharp bound with parameters $(\tilde\Lambda,\tilde\Gamma)$
lies between the corresponding deMSM sharp bound with $\Lambda=\Gamma=\min\{\tilde\Lambda,\tilde\Gamma\}$ and that with $\Lambda=\Gamma=\max\{\tilde\Lambda,\tilde\Gamma\}$.
With the choice $\Lambda=\Gamma$, the two constraints \eqref{eq:msm} and \eqref{eq:demsm} appear, a priori, as plausible as each other,
but deMSM sharp bounds are always tighter than the MSM ones with $\Lambda_1^{-1}=\Lambda_2=\Lambda$ in \eqref{eq:msm-2006} and the MSM-U ones with constraint \eqref{eq:msm} only.

The preceding interpretation of $(\Lambda,\Gamma)$ through quantile and shrinkage levels
is data-independent, i.e., the meanings of fixed choices of $(\Lambda,\Gamma)$ are the same across different studies.
The resulting sensitivity bounds are, however, data-dependent. In particular, the deviation \eqref{eq:deMSM-deviation} may
differ from study to study even for the same choices of $(\Lambda,\Gamma)$.
For the suggested choice $\Lambda=\Gamma$, sensitivity analysis using deMSM can be conducted by selecting a grid of values for $\Lambda=\Gamma$ data-independently,
and evaluating the corresponding sensitivity bounds which are data-dependent.
The greater the value of $\Lambda=\Gamma$ that reverses the causal conclusion from under the unconfoundedness assumption,
the more robust the study is to unmeasured confounding.

Our discussion so far deals with population sensitivity bounds. With sample data, the population bounds need to be estimated, together with confidence intervals
to account for sampling uncertainty. Fortunately, as shown by Proposition \ref{pro:demsm-mu1-up},
population sharp bounds from deMSM coincide with the MSM sharp bounds with implied parameters $(\bar\Lambda_1,\bar\Lambda_2)$.
Therefore, the estimation method developed for MSM population bounds \citep{tan2024model} can be directly adopted for estimating deMSM population bounds, including
doubly robust point estimation and confidence intervals.

%% file: Sec4Implication.tex
The development of deMSM also has various implications for eMSM.
First, deMSM differs from eMSM only in the outcome sensitivity constraint. In fact, the deMSM distributional constraint \eqref{eq:demsm} implies eMSM mean constraint \eqref{eq:emsm} with the \emph{implied} outcome sensitivity parameters
\begin{align}\label{eq:implied-eMSM}
    \hspace{-1.5em}\bar\Delta_1(X)=\{\Gamma_2(X)-\Gamma_1(X)\}E\left(\rho_{1,1-\tau_{\scaleto{\Gamma}{3pt}}}^*|T=1,X\right),\bar\Delta_2(X)=\{\Gamma_2(X)-\Gamma_1(X)\}E\left(\rho_{1,\tau_{\scaleto{\Gamma}{3pt}}}^*|T=1,X\right).
\end{align}
By comparing the expressions \eqref{eq:eMSM-mu1-upper} with implied $(\bar\Delta_1,\bar\Delta_2)$ and \eqref{eq:nu1+-min},
the sharp bounds of deMSM are in general tighter, sometimes strictly tighter, than those of the implied eMSM.

Second, while it may not be feasible to match deMSM sharp bounds with eMSM ones, the eMSM sharp bounds can be matched with deMSM ones using proper $(\Gamma_1,\Gamma_2)$. We consider here eMSM sharp bounds under the recommended specification from \citet{zhang2025enhanced}, where $(\Lambda_1,\Lambda_2)$ are constant and $(\Delta_1,\Delta_2)$ are re-parameterized with a constant $\delta\in[0,1]$ as
\begin{align}\label{eq:emsm-recspec}
\begin{split}
& \Delta_1(X)= \delta /(1-\max\{1-\tau,\tau\}) \cdot E\left( \rho_{1,1-\max\{1-\tau,\tau\}}^*|T=1,X\right),\\
& \Delta_2(X)= \delta /(1-\max\{1-\tau,\tau\}) \cdot E\left( \rho_{1,\max\{1-\tau,\tau\}}^*|T=1,X\right),
\end{split}
\end{align}
For simplicity, we assume $\tau\ge1/2$, which holds for $\Lambda_1^{-1}=\Lambda_2=\Lambda$. In this case, the eMSM sharp bounds of $\nu^1$ are simplified to
\begin{align}\label{eq:emsm-sharp-rec}
\begin{split}
    &\nu^{1+}_{\mytext{eMSM}} (X)=E(Y|T=1,X)+\delta(\Lambda_2-\Lambda_1)E\left(\rho_{1,\tau}^*|T=1,X \right),\\
    &\nu^{1-}_{\mytext{eMSM}} (X)=E(Y|T=1,X)-\delta(\Lambda_2-\Lambda_1)E\left(\rho_{1,1-\tau}^*|T=1,X \right).
\end{split}
\end{align}
By comparing \eqref{eq:demsm-nu1-up} and \eqref{eq:emsm-sharp-rec}, we find that, for a given $\delta$, the eMSM sharp bounds of $\nu^1(X)$
are simultaneously matched by the corresponding deMSM sharp bounds with
\begin{align} \label{eq:emsm-implied-gam}
  \Gamma_1=1-\delta,\quad \Gamma_2=1+\delta\odds(\tau),
\end{align}
satisfying $\tauG=\tau$,
where $\odds(c)=c/(1-c)$. The specification of $(\Gamma_1,\Gamma_2)$ is not unique for deMSM to match the eMSM sharp bounds. In fact, $\Gamma_1=1-\delta$ with any $\Gamma_2\ge1+\delta\odds(\tau)$ also achieves the matching,
and hence the choice \eqref{eq:emsm-implied-gam} is optimal in taking the smallest $\Gamma_2$.
Moreover, for the choice \eqref{eq:emsm-implied-gam}, the deMSM implied sensitivity parameters $(\bar\Delta_1,\bar\Delta_2)$ in \eqref{eq:implied-eMSM} match the eMSM outcome sensitivity parameters $(\Delta_1,\Delta_2)$ in \eqref{eq:emsm-recspec}. See Supplement Section \ref{sec:proof-implied-eMSM} for details. The simple choice 
$\Gamma_2^{-1}=\Gamma_1=1-\delta$ also achieves the matching if $\delta\ge (2\tau-1)/\tau$ (i.e., $1/(1-\delta) \ge1+\delta\odds(\tau)$), while it leads to narrower sharp bounds than those of eMSM with the given $\delta$ if $\delta< (2\tau-1)/\tau$.
But the implied $(\bar\Delta_1,\bar\Delta_2)$ by this choice differ from $(\Delta_1,\Delta_2)$ in \eqref{eq:emsm-recspec}.
Together these matchings of the eMSM and deMSM sharp bounds indicate that the eMSM parameter $\delta$ is aligned with $1-\Gamma_1$ in deMSM.


Lastly, inspired by the alignment between eMSM and deMSM, we see that an eMSM specified by $(\Lambda_1,\Lambda_2,\delta)$ can also be aligned with an MSM such that the eMSM sharp bounds in \eqref{eq:emsm-sharp-rec} are matched by MSM ones in \eqref{eq:msm-sharp}. The resulting MSM is specified with parameters
\begin{align}\label{eq:emsm-implied-lam}
    \tilde\Lambda_1=1-\delta(1-\Lambda_1) , \quad \tilde\Lambda_2=1+\delta(\Lambda_2-1).
\end{align}
The expressions in \eqref{eq:emsm-implied-lam} shows that the eMSM outcome sensitivity constraint, when re-parameterized by $\delta$, has a shrinkage effect on the treatment sensitivity parameters $(\Lambda_1,\Lambda_2)$, i.e., the deviations from no unmeasured confounding coded by $1-\Lambda_1$ and $\Lambda_2-1$ are shrunk proportionally to $\delta$. A caveat is that $(\tilde\Lambda_1,\tilde\Lambda_2)$ do not satisfy the simple relation $\tilde\Lambda_1=\tilde\Lambda_2^{-1}$.

%% file: supplementary.tex
\setcounter{page}{1}

\setcounter{section}{0}
\setcounter{equation}{0}

\setcounter{table}{0}
\setcounter{figure}{0}

\setcounter{pro}{0}
\renewcommand{\thepro}{S\arabic{pro}}

\setcounter{lem}{0}
\renewcommand{\thelem}{S\arabic{lem}}

\setcounter{thm}{0}
\renewcommand{\thethm}{S\arabic{thm}}

\setcounter{ass}{0}
\renewcommand{\theass}{S\arabic{ass}}

\renewcommand\thesection{\Roman{section}}

\renewcommand\thesubsection{\thesection.\arabic{subsection}}

\renewcommand\theequation{S\arabic{equation}}

\renewcommand\thetable{S\arabic{table}}
\renewcommand\thefigure{S\arabic{figure}}

\if1\blind
{
  \begin{center}
{\Large Supplementary Material for}\\
{\Large``Distributionally Enhanced Marginal Sensitivity Model and Bounds''}

\vspace{.1in}
Yi Zhang, Wenfu Xu, Zhiqiang Tan
\vspace{.1in}
\end{center}
} \fi
\section{Additional deMSM bounds}\label{sec:deMSM-bounds}
In the main text, the deMSM model and related results are developed for $\mu^1$. We extend the discussion to $\mu^0$ and ATE.

Similar to \eqref{eq:demsm}, the deMSM model for $\mu^0$ is defined as
\begin{align}\label{eq:demsm-Y0}
    \omega^*_0(Y^0,X,U)=\frac{\ud P(Y^0|X,U)}{\ud P(Y^0|T=0,X)}\in[\Gamma_1^\prime(X),\Gamma_2^\prime(X)],
\end{align}
where the sensitivity parameters $(\Gamma_1^\prime,\Gamma_2^\prime)$ are generally different from $(\Gamma_1,\Gamma_2)$. We refer to the model defined by \eqref{eq:unm-conf-indep}, \eqref{eq:msm}, \eqref{eq:demsm} and \eqref{eq:demsm-Y0} as complete deMSM, and the model defined by\eqref{eq:unm-conf-indep}, \eqref{eq:msm} and either \eqref{eq:demsm} or \eqref{eq:demsm-Y0} as (partial) deMSM.

By symmetry of $T$ in its two labels and Propositions \ref{pro:demsm-mu1-up}, the sharp lower bound and upper bound of $\nu^{0}(X)=E(Y^0|T=1,X)$ under the partial deMSM defined by \eqref{eq:unm-conf-indep}, \eqref{eq:msm} and \eqref{eq:demsm-Y0} are respectively
\begin{align}
\begin{split}
    &\nu^{0-}_{\mytext{deMSM}}(X) = E(Y|T=0,X)-\left\{\bar\Lambda_2^\prime(X)-\bar\Lambda_1^\prime(X)\right\}E \left(\rho_{0,1-\bar\tau^\prime}^*|T=0,X \right),\\
&\nu^{0+}_{\mytext{deMSM}}(X) = E(Y|T=0,X)+\left\{\bar\Lambda_2^\prime(X)-\bar\Lambda_1^\prime(X)\right\}E \left(\rho_{0,\bar\tau^\prime}^*|T=0,X \right),\label{eq:demsm-nu0}
\end{split}
\end{align}
where $\bar\tau'(X)$ is defined the same as $\bar\tau(X)$ except with  $(\bar\Lambda_1^\prime,\bar\Lambda_2^\prime)$ replacing $(\bar\Lambda_1,\bar\Lambda_2)$,  $\rho_{0,\gamma}^*=\rho_\gamma(Y,q_{0,\gamma}^*)$ and $q_{0,\gamma}^*(X)$ is the $\gamma(X)$-quantile of $Y|T=0,X$ for $\gamma\in(0,1)$, and
\begin{align}
&\bar\Lambda_1^\prime(X)=1 -\{\Lambda_1^{-1}(X)-\Lambda_2^{-1}(X)\}\{\Gamma_2^\prime(X)-\Gamma_1^\prime(X)\} \min\big[ \{1-\tau^\prime(X)\}\{1-\tau_{\scaleto{\Gamma}{4pt}}^\prime(X)\}, \tau^\prime(X)\tau_{\scaleto{\Gamma}{4pt}}^\prime(X) \big],\\
    &\bar\Lambda_2^\prime(X)=1 +\{\Lambda_1^{-1}(X)-\Lambda_2^{-1}(X)\}\{\Gamma_2^\prime(X)-\Gamma_1^\prime(X)\}  \min\big[ \{1-\tau^\prime(X)\}\tau_{\scaleto{\Gamma}{4pt}}^\prime(X), \tau^\prime(X)\{1-\tau_{\scaleto{\Gamma}{4pt}}^\prime(X)\} \big],
\end{align}
with $\tau^\prime(X)$ and $\tauG^\prime(X)$ defined by substituting $(\Lambda_2^{-1},\Lambda_1^{-1})$ and $(\Gamma_1^\prime,\Gamma_2^\prime)$ respectively for $(\Lambda_1,\Lambda_2)$ in $\tau(X)$.
The sharp bounds on $\mu^{0}$ are
\begin{align}
    \mu^{0+}_{\rm{deMSM}} = E \{(1-T)Y+T\nu^{0+}_{\rm{deMSM}} (X) \},\quad \mu^{0-}_{\rm{deMSM}} = E \{(1-T)Y+T\nu^{0-}_{\rm{deMSM}} (X) \}.
\end{align}

The sharp bounds of $\nu^0(X)$ and $\nu^1(X)$, hence those of $\mu^0$ and $\mu^1$, under the complete deMSM are identical to those given by the partial deMSMs. Moreover, as proved in Supplement Section \ref{sec:proof-prop1}, the sharp lower bound of $\mu^0$ and the sharp upper bound of $\mu^1$ under respective partial deMSMs (or the sharp upper bound of $\mu^0$ and the sharp lower bound of $\mu^1$) are attained simultaneously by a single distribution $Q$ allowed in the complete deMSM. Due to this simultaneous sharpness, the sharp bounds of a contrast of $\mu^0$ and $\mu^1$, such as ATE, are obtained by contrasting the sharp bounds of $\mu^0$ and $\mu^1$. For example, the sharp upper bound and sharp lower bound of ATE under the complete deMSM are $\mu^{1+}_{\rm{deMSM}}-\mu^{0-}_{\rm{deMSM}}$ and $\mu^{1-}_{\rm{deMSM}}-\mu^{0+}_{\rm{deMSM}}$ respectively.

\section{Proofs of main results}\label{sec:proof-main}
\subsection{Proofs related to the implied MSM sensitivity parameters}\label{sec:proof-implied-msm2006}
\noindent\textbf{Proof of \eqref{eq:implied-msm2006}.}
To start, we note
\begin{align}
    \lambda^*_1(Y^1=y,X)&=\int \omega^*(y,X,u)\lambda^*(X,u)\ud P(U=u|T=1)\\
    &\le\sup_Q\int \omega_Q(y,X,u)\lambda_Q(X,u)\ud Q(U=u|T=1),\label{eq:lambda1-demsm}
\end{align}
where the optimization is over $Q$ in the deMSM specified by \eqref{eq:unm-conf-indep}, \eqref{eq:msm} and \eqref{eq:demsm}, i.e., $Q$ such that:
\begin{itemize}
\vspace{-.1in}
\addtolength{\itemsep}{-.1in}
\item[(i)] $Q$ and $P$ induce the same distribution of $(Y,T,X)$ under the consistency assumption;
\item[(ii)] \eqref{eq:unm-conf-indep}, \eqref{eq:msm} and \eqref{eq:demsm} hold under $Q$.
\vspace{-.1in}
\end{itemize}
For any $y$, $\omega_Q(y,X,U)$ is a measurable function of $X$ and $U$ that satisfies $E_Q\{\omega_Q(y,X,U)|T=1,X\}=1$ and $\omega_Q(y,X,U)\in[\Gamma_1(X),\Gamma_2(X)]$. Similarly, $\lambda_Q(X,U)$ is a measurable function of $X$ and $U$ that satisfies $E_Q\{\lambda_Q(X,U)|T=1,X\}=1$ and $\lambda_Q(X,U)\in[\Lambda_1(X),\Lambda_2(X)]$. The converse is also true. For any nonnegative measurable functions $\omega(\cdot)$ and $\lambda(\cdot)$ of $X$ and $U$ satisfying the sensitivity constraints
\begin{align}\label{eq:sens-constraint}
    \lambda(X,U)\in [\Lambda_1(X),\Lambda_2(X)], \quad \omega(X,U)-1\in[-(1-\Gamma_1(X)),\Gamma_2(X)-1],
\end{align}
and inherent constraints
\begin{align}\label{eq:inh-constraint}
\int \lambda(X,u) \ud Q(U=u|T=1,X)=1,\quad\int \omega(X,u)\ud Q(U=u|T=1,X)=1
\end{align}
there exists a joint distribution $Q$ satisfying conditions (i) and (ii) so that $\omega_Q(y,X,U)=\omega(X,U)$ and $\lambda_Q(X,U)=\lambda(X,U)$. Consequently, \eqref{eq:lambda1-demsm} is equivalent to the optimization over all conditional distribution $Q(U|T=1,X)$ and nonnegative measurable function $(\omega(y,\cdot), \lambda(\cdot))$:
\begin{align}
    \sup_{Q(U|T=1,X),\omega,\lambda}\int \omega(y,X,u)\lambda(X,u)\ud Q(U=u|T=1,X)\label{eq:opt-v_e}
\end{align}
subject to constraints \eqref{eq:sens-constraint} and \eqref{eq:inh-constraint}. This problem corresponds to a special case, with $\eta(\cdot)=\omega(\cdot)$, of the optimization problem solved in \citeappend{zhang2025enhanced} using a generalized Neyman--Pearson lemma. The solution gives $\bar\Lambda_2(X)$ in \eqref{eq:implied-msm2006}. The lower bound in \eqref{eq:implied-msm2006} comes from the optimization that replaces supremum with infimum in \eqref{eq:opt-v_e}.

\noindent\textbf{Proof of \eqref{eq:impliedpara-demsmpara}.}
By the definition of $\tau$ and that of $\tauG$,
\begin{align}
\bar\Lambda_1(X)&=1 -\min\big[ \{1-\Lambda_1(X)\}\{1-\Gamma_1(X)\}, \{\Lambda_2(X)-1\}\{\Gamma_2(X)-1\} \big]\\
&\ge 1-\{1-\Lambda_1(X)\}\{1-\Gamma_1(X)\}\\
&\ge 1- \min\{1-\Lambda_1(X),1-\Gamma_1(X)\}=\max\{\Lambda_1(X),\Gamma_1(X)\},\\
\bar\Lambda_2(X)&=1 + \min\big[ \{1-\Lambda_1(X)\}\{\Gamma_2(X)-1\}, \{\Lambda_2(X)-1\}\{1-\Gamma_1(X)\} \big]\\
&\le1+\min \{\Gamma_2(X)-1, \Lambda_2(X)-1\}=\min\{\Gamma_2(X),\Lambda_2(X)\}.
\end{align}

\subsection{Proof of Proposition \ref{pro:demsm-mu1-up}}\label{sec:proof-prop1}
It is established in Section \ref{sec:deMSM} that \eqref{eq:demsm-nu1-up} is a valid upper bound for $\nu^{1}(X)$ under deMSM. We complete the proof by constructing a joint distribution $Q$ that not only belongs to the partial deMSMs, but also belongs to the complete deMSM, i.e.,
\begin{itemize}\addtolength{\itemsep}{-.1in}
\item[(i)] $Q$ and $P$ induce the same distribution of $(Y,T,X)$ under the consistency assumption;
\item[(ii)] \eqref{eq:unm-conf-indep}, \eqref{eq:msm}, \eqref{eq:demsm} and \eqref{eq:demsm-Y0} hold under $Q$,
\end{itemize}
and that $E_Q(Y^1)=\mu^{1+}_{\rm{deMSM}}$ and $E_Q(Y^0)=\mu^{0-}_{\rm{deMSM}}$.

For simplicity of presentation, we suppress $X$ in all notation. The construction of $Q$ is conditional on $X$. We assume $\ud Q(X)=\ud P(X)$, where $\ud P$ and $\ud Q$ are density functions with respect to some dominating meausre.

\noindent\textbf{Construction of $Q$.} To start, we define $Q(T)=P(T)$, and let the unmeasured confounder $U$ be binary such that
\begin{align}
&Q (U = 1 | T=1) = 1-Q(U = 0 | T=1) = 1-\tau,\label{eq:Q-U|T=1}\\
&Q(U = 1 | T=0)=\tau^\prime=(\Lambda_1^{-1}-1)/(\Lambda_1^{-1}-\Lambda_2^{-1})=\Lambda_2(1-\tau). \label{eq:Q-U|T=0}
\end{align}
Define $\ud Q(Y^1=y|T,U=1)$ and $\ud Q(Y^1=y|T,U=0)$ through
\begin{align}
    &\frac{\ud Q(Y^1=y|T,U=1)}{\ud P(Y=y|T=1)}=\begin{cases}\label{eq:Q-Y^1|U=1}
        \bar\Gamma_1&\quad\text{if}\quad y<q^*_{1,\bar\tau}\\
        \bar\Gamma_2&\quad\text{if}\quad y>q^*_{1,\bar\tau}\\
        \begin{cases}
            \bar\Gamma_2 \quad\text{w.p.}\quad \frac{P(Y\le q^*_{1,\bar\tau}|T=1)-\tau}{P(Y=q^*_{1,\bar\tau}|T=1)}\\
            \bar\Gamma_1 \quad\text{w.p.}\quad\frac{\tau-P(Y< q^*_{1,\bar\tau}|T=1)}{P(Y=q^*_{1,\bar\tau}|T=1)}
        \end{cases} &\quad\text{if}\quad y=q^*_{1,\bar\tau}
    \end{cases},\\
   &\tau\ud Q(Y^1=y|T,U=0)+(1-\tau)\ud Q(Y^1=y|T,U=1)=\ud P(Y=y|T=1). \label{eq:Q-Y^1|U=0}
\end{align}
where $\bar\Gamma_1=\max\{\Gamma_1,1+\odds (\tau)(1-\Gamma_2)\}$ and $\bar\Gamma_2=\min\{\Gamma_2,1+\odds (\tau)(1-\Gamma_1)\}$. Similarly, define $\ud Q(Y^0=y|T,U=1)$ and $\ud Q(Y^0=y|T,U=0)$ through
\begin{align}
    &\frac{\ud Q(Y^0=y|T,U=1)}{\ud P(Y=y|T=0)}=\begin{cases}\label{eq:Q-Y^0|U=1}
        \bar\Gamma_1^\prime&\quad\text{if}\quad y<q^*_{0,1-\bar\tau^\prime}\\
        \bar\Gamma_2^\prime&\quad\text{if}\quad y>q^*_{0,1-\bar\tau^\prime}\\
        \begin{cases}
            \bar\Gamma_2^\prime \quad\text{w.p.}\quad \frac{P(Y\le q^*_{0,1-\bar\tau^\prime}|T=0)-1+\tau^\prime}{P(Y=q^*_{0,1-\bar\tau^\prime}|T=0)}\\
            \bar\Gamma_1^\prime \quad\text{w.p.}\quad\frac{1-\tau^\prime-P(Y< q^*_{0,1-\bar\tau^\prime}|T=0)}{P(Y=q^*_{0,1-\bar\tau^\prime}|T=0)}
        \end{cases} &\quad\text{if}\quad y=q^*_{0,1-\bar\tau^\prime}
    \end{cases},\\
   &(1-\tau^\prime)\ud Q(Y^0=y|T,U=0)+\tau^\prime\ud Q(Y^0=y|T,U=1)=\ud P(Y=y|T=0). \label{eq:Q-Y^0|U=0}
\end{align}
where $\bar\Gamma_1^\prime=\max\{\Gamma_1^\prime,1+\odds (1-\tau^\prime)(1-\Gamma_2^\prime)\}$ and $\bar\Gamma_2^\prime=\min\{\Gamma_2^\prime,1+\odds (1-\tau^\prime)(1-\Gamma_1^\prime)\}$.
Finally, we define
\begin{align}\label{eq:Q-Y0Y1}
  \ud Q(Y^0,Y^1|T,U)=\ud Q(Y^0|T,U)\ud Q(Y^1|T,U).
\end{align}

\noindent\textbf{Proof of (i) and (ii).} By \eqref{eq:Q-U|T=1} and \eqref{eq:Q-U|T=0}, $\lambda_Q(X,U=1)=\Lambda_2$. Similarly, $\lambda_Q(X,U=0)=(1-\tau^\prime)/\tau=\Lambda_1$. Therefore, \eqref{eq:msm} holds under $Q$.

By \eqref{eq:Q-Y^1|U=1}, $\omega_Q(Y^1,X,U=1)\in[\bar\Gamma_1,\bar\Gamma_2]\subseteq[\Gamma_1,\Gamma_2]$, which together with \eqref{eq:Q-Y^1|U=0} shows that
\begin{align}
  \omega_Q(Y^1,X,U=0)&=1/\tau-\odds(1-\tau)\omega_Q(Y^1,X,U=1)\\
  &\in[1/\tau-\odds(1-\tau)-1+\Gamma_1,1/\tau-\odds(1-\tau)-1+\Gamma_2]=[\Gamma_1,\Gamma_2].
\end{align}
Therefore, \eqref{eq:demsm} holds under $Q$. Moreover, \eqref{eq:Q-Y^1|U=0} is equivalent to
\begin{align*}
\ud  Q(Y^1 =y | T=1) & =Q(U=0|T=1)\ud Q(Y^1=y|T=1,U=0)\\
&+Q(U=1|T=1)\ud Q(Y^1=y|T=1,U=1)\\
& =\ud P(Y=y|T=1),
\end{align*}
which proves that under the consistency assumption, $\ud Q(Y|T=1)=\ud  Q(Y^1 | T=1)=\ud P(Y|T=1)$. As $Q(T)=P(T)$, the induced distributions of $Q$ and $P$ agree on $(Y,T=1)$.
With similar arguments, \eqref{eq:Q-Y^0|U=1} and \eqref{eq:Q-Y^0|U=0} establish \eqref{eq:demsm-Y0} under $Q$ and that the induced distributions of $Q$ and $P$ agree on $(Y,T=0)$ under the consistency assumption.
This completes the proof for (i).

By construction, we have $Y^t\perp T|U$ for $t=0,1$. With \eqref{eq:Q-Y0Y1}, $\ud Q(Y^0,Y^1|T=1,U)=\ud Q(Y^0|T=0,U)\ud Q(Y^1|T=0,U)=\ud Q(Y^0,Y^1|T=0,U)$, which proves \eqref{eq:unm-conf-indep} and completes the proof for (ii).

\noindent\textbf{Proof of $E_Q(Y^1)=\mu^{1+}_{\rm{deMSM}}$ and $E_Q(Y^0)=\mu^{0-}_{\rm{deMSM}}$.} According to \citeappend{tan2024model}, a density ratio $\lambda_{1,Q^+}(y)=\ud Q^+(Y^1=y|T=0)/\ud P(Y=y|T=1)$ (satisfying the normalization condition)
achieves the MSM sharp upper bound of $\mu^1$ with parameters $(\bar\Lambda_1,\bar\Lambda_2)$ if
\begin{align}\label{eq:msm-lam1+}
    \lambda_{1,Q^+}(y)=\begin{cases}
        \bar\Lambda_1 &\quad\text{if}\quad y<q^*_{1,\bar\tau}\\
        \bar\Lambda_2 &\quad\text{if}\quad y>q^*_{1,\bar\tau}
    \end{cases},
\end{align}
and a density ratio $\lambda_{0,Q^-}(y)=\ud Q^-(Y^0=y|T=1)/\ud P(Y=y|T=0)$ (satisfying the normalization condition)
achieves the MSM sharp lower bound of $\mu^0$ with parameters $(\bar\Lambda_1^\prime,\bar\Lambda_2^\prime)$ if
\begin{align}\label{eq:msm-lam0-}
    \lambda_{0,Q^-}(y)=\begin{cases}
        \bar\Lambda_1^\prime &\quad\text{if}\quad y>q^*_{0,1-\bar\tau^\prime}\\
        \bar\Lambda_2^\prime &\quad\text{if}\quad y<q^*_{0,1-\bar\tau^\prime}
    \end{cases}.
\end{align}
It suffices to show that $Q$ constructed above satisfies \eqref{eq:msm-lam1+} and \eqref{eq:msm-lam0-} with $Q$ replacing $Q^+$ and $Q^-$.

For the $Q$ constructed above,
\begin{align}
    \lambda_{1,Q}(y)&=\frac{\tau^\prime\ud Q(Y^1=y|T=0,U=1)+(1-\tau^\prime)\ud Q(Y^1=y|T=0,U=0)}{\ud P(Y=y|T=1)}\\
    &=\frac{(1-\Lambda_1)Q(Y^1=y|T=0,U=1)+\Lambda_1\ud P(Y=y|T=1)}{\ud P(Y=y|T=1)}\\
    &=\begin{cases}
        (1-\Lambda_1)\bar\Gamma_1+\Lambda_1&\quad\text{if}\quad y<q^*_{1,\bar\tau}\\
         (1-\Lambda_1)\bar\Gamma_2+\Lambda_1&\quad\text{if}\quad y>q^*_{1,\bar\tau}\\
    \end{cases}\\
    &=\begin{cases}
        1-\min\{(1-\Lambda_1)(1-\Gamma_1),(\Lambda_2-1)(\Gamma_2-1)\}=\bar\Lambda_1&\quad\text{if}\quad y<q^*_{1,\bar\tau}\\
        1+\min\{(1-\Lambda_1)(\Gamma_2-1),(\Lambda_2-1)(1-\Gamma_2)\}=\bar\Lambda_2&\quad\text{if}\quad y>q^*_{1,\bar\tau}
    \end{cases},
\end{align}
where we used iterated expectations in the first line, \eqref{eq:Q-Y^1|U=0} in the second line, and \eqref{eq:Q-Y^1|U=1} in the third line. Similarly,
\begin{align}
    \lambda_{0,Q}(y)
    &=\begin{cases}
        -(\Lambda_1^{-1}-1)\bar\Gamma_1^\prime+\Lambda_1^{-1}&\quad\text{if}\quad y<q^*_{0,1-\bar\tau^\prime}\\
        -(\Lambda_1^{-1}-1)\bar\Gamma_2^\prime+\Lambda_1^{-1}&\quad\text{if}\quad y>q^*_{0,1-\bar\tau^\prime}
    \end{cases}\\
    &=\begin{cases}
        1+\min\{(\Lambda_1^{-1}-1)(1-\Gamma_1^\prime),(1-\Lambda_2^{-1})(\Gamma_2^\prime-1)\}=\bar\Lambda_2^\prime&\quad\text{if}\quad y<q^*_{0,1-\bar\tau^\prime}\\
        1-\min\{(1-\Lambda_2^{-1})(1-\Gamma_1^\prime),(\Lambda_1^{-1}-1)(\Gamma_2^\prime-1)\}=\bar\Lambda_1^\prime&\quad\text{if}\quad y>q^*_{0,1-\bar\tau^\prime}
    \end{cases}.
\end{align}

\subsection{Proofs related to eMSM}\label{sec:proof-implied-eMSM}
\noindent\textbf{Proof of \eqref{eq:implied-eMSM}.}
To start, we have
\begin{align}
    \eta^*(X,u)&=\int y \omega^*(y,X,u)\ud P(Y^1=y|T=1,X),\label{eq:eta-omega}\\
    &\le\sup_{\omega}\int y \omega(y,X)\ud P(Y^1=y|T=1,X),
\end{align}
where, for any $u$, the optimization is over every $\omega(\cdot,u)$, a measurable function of $Y^1$ and $X$, subject to $E\{\omega(Y^1,X,u)|T=1,X\}=1$ and $\omega(Y^1,X,u)\in[\Gamma_1(X),\Gamma_2(X)]$. Such an optimization also defines the sharp upper bound of $\nu^1(X)$ under the original MSM model \eqref{eq:msm-2006} \citepappend{tan2024model}, and hence admits the solution \eqref{eq:eMSM-mu1-upper} with $(\Gamma_1(X),\Gamma_2(X))$ replacing $(\Lambda_1(X),\Lambda_2(X))$, which gives $\Delta_2$ as in \eqref{eq:implied-eMSM}. The argument for $\Delta_1(X)$ is similar.

\noindent\textbf{Proof of equivalence between \eqref{eq:implied-eMSM} and \eqref{eq:emsm-recspec} under specification \eqref{eq:emsm-implied-gam}.}

With $(\Gamma_1=1-\delta,\Gamma_2=1+\delta\odds(\tau))$ and $\tau\ge1/2$, $\tauG=\tau$ and hence $\bar\tau=\tau$. The implied eMSM outcome sensitivity parameters are 
\begin{align}
 \bar\Delta_1(X)=\delta/(1-\tau)E\left(\rho_{1,1-\tau}^*|T=1,X\right),\bar\Delta_2(X)=\delta/(1-\tau)E\left(\rho_{1,\tau}^*|T=1,X\right),
\end{align}
which matches the recommended $(\Delta_1,\Delta_2)$--specification in \eqref{eq:emsm-recspec}.

\noindent\textbf{Proof of matched sharp bounds between deMSM and eMSM under specification \eqref{eq:emsm-implied-gam}.}

Following the proceeding proof, the implied MSM sensitivity parameters are
\begin{align}
  \bar\Lambda_1&=1 -\min \{(1-\Lambda_1)(1-\Gamma_1), (\Lambda_2-1)(\Gamma_2-1)\}=1-\delta(1-\Lambda_1),\\
  \bar\Lambda_2&=1+\min\{(1-\Lambda_1)(\Gamma_2-1), (\Lambda_2-1)(1-\Gamma_1)\}=1+\delta(\Lambda_2-1).
\end{align}
Therefore, by Proposition \ref{pro:demsm-mu1-up},
\begin{align}
&\nu^{1-}_{\mytext{deMSM}}(X) = E(Y|T=1,X)-\delta\left(\Lambda_2-\Lambda_1\right)E \left(\rho_{1,1-\tau}^*|T=1,X \right),\\
&\nu^{1+}_{\mytext{deMSM}}(X) = E(Y|T=1,X)+\delta\left(\Lambda_2-\Lambda_1\right)E \left(\rho_{1,\tau}^*|T=1,X \right),
\end{align}
which coincide with the eMSM sharp bounds in \eqref{eq:emsm-sharp-rec}.

\noindent\textbf{Proof of eMSM sharp bounds \eqref{eq:emsm-sharp-rec} matched by \eqref{eq:demsm-nu1-up} in general.}
To match \eqref{eq:emsm-sharp-rec} with \eqref{eq:demsm-nu1-up} for arbitrary data and $(\Lambda_1,\Lambda_2)$ such that $\tau\ge1/2$, a sufficient and necessary condition is that $\bar\tau=\tau$ and
\begin{align}
   (\Gamma_2-\Gamma_1)\big[ \min\left\{(1-\tau)(1-\tauG), \tau\tauG\right\}+\min\left\{(1-\tau)\tauG, \tau(1-\tauG)\right\}\big]=\delta,
\end{align}
which happens if and only if $\tauG\ge\tau$ and $\delta=1-\Gamma_1$. The condition $\tauG\ge\tau$ then leads to $\Gamma_2\ge1+\delta\odds(\tau).$

\noindent\textbf{Proof of matched sharp bounds between deMSM and eMSM with symmetric $(\Gamma_1,\Gamma_2)$.}
From previous proof, $\Gamma_2=\Gamma_1^{-1}=1/(1-\delta)$ is feasible if and only if $1/(1-\delta)\ge1+\delta\odds(\tau)$. Equivalently, $\delta\ge(2\tau-1)/\tau$ as claimed in Section \ref{sec:implication}. 

%% file: main.bbl
\begin{thebibliography}{}

\bibitem[Tan, 2024]{tan2024model}
Tan, Z. (2024).
\newblock Model-assisted sensitivity analysis for treatment effects under unmeasured confounding via regularized calibrated estimation.
\newblock {\em Journal of the Royal Statistical Society Ser. B}, 86:1339–1363.

\bibitem[Zhang et~al., 2025]{zhang2025enhanced}
Zhang, Y., Xu, W., and Tan, Z. (2025).
\newblock Enhanced marginal sensitivity model and bounds.
\newblock {\em arXiv preprint arXiv:2504.08301}.

\end{thebibliography}


\begin{thebibliography}{}

\bibitem[Dorn and Guo, 2023]{dorn2023sharp}
Dorn, J. and Guo, K. (2023).
\newblock Sharp sensitivity analysis for inverse propensity weighting via
  quantile balancing.
\newblock {\em Journal of the American Statistical Association},
  118:2645--2657.

\bibitem[Greenland and Robins, 1986]{greenland1986identifiability}
Greenland, S. and Robins, J.~M. (1986).
\newblock Identifiability, exchangeability, and epidemiological confounding.
\newblock {\em International Journal of Epidemiology}, 15:413--419.

\bibitem[Koenker and Bassett, 1978]{koenker1978regression}
Koenker, R. and Bassett, Jr., G. (1978).
\newblock Regression quantiles.
\newblock {\em Econometrica}, 46:33--50.

\bibitem[Rosenbaum, 2002]{PRosenbaum2002}
Rosenbaum, P.~R. (2002).
\newblock {\em Observational studies \rm{(2nd ed.)}}.
\newblock Springer.

\bibitem[Rubin, 1974]{rubin1974estimating}
Rubin, D.~B. (1974).
\newblock Estimating causal effects of treatments in randomized and
  nonrandomized studies.
\newblock {\em Journal of Educational Psychology}, 66:688.

\bibitem[Splawa-Neyman, 1990]{Neyman1923}
Splawa-Neyman, J. (1990).
\newblock On the application of probability theory to agricultural experiments:
  Essay on principles, {Section} 9.
\newblock {\em Statistical Science}, 5:465--472.

\bibitem[Tan, 2006]{tan2006distributional}
Tan, Z. (2006).
\newblock A distributional approach for causal inference using propensity
  scores.
\newblock {\em Journal of the American Statistical Association},
  101:1619--1637.

\bibitem[Tan, 2024]{tan2024model}
Tan, Z. (2024).
\newblock Model-assisted sensitivity analysis for treatment effects under
  unmeasured confounding via regularized calibrated estimation.
\newblock {\em Journal of the Royal Statistical Society Ser. B},
  86:1339–1363.

\bibitem[Zhang et~al., 2025]{zhang2025enhanced}
Zhang, Y., Xu, W., and Tan, Z. (2025).
\newblock Enhanced marginal sensitivity model and bounds.
\newblock {\em arXiv preprint arXiv:2504.08301}.

\end{thebibliography}
